\documentclass[conference]{IEEEtran}
\usepackage{CJKutf8}
\IEEEoverridecommandlockouts
\usepackage{cite}
\usepackage{amsmath,amssymb,amsfonts}
\usepackage{algorithmic}
\usepackage{graphicx}
\usepackage{textcomp}
\usepackage{xcolor}
\usepackage[table]{xcolor}
\usepackage{transparent}
\usepackage{diagbox}
\usepackage{multirow}
\usepackage{bm}
\usepackage{float} 
\usepackage{subcaption}
\usepackage{booktabs}
\usepackage{balance}
\usepackage{pifont}

\usepackage[hidelinks]{hyperref}
\usepackage{cleveref}
\usepackage[ruled,vlined,linesnumbered]{algorithm2e} 

\usepackage{microtype}          
\def\BibTeX{{\rm B\kern-.05em{\sc i\kern-.025em b}\kern-.08em
    T\kern-.1667em\lower.7ex\hbox{E}\kern-.125emX}}
    
\begin{document}
\begin{CJK*}{UTF8}{gbsn}
\title{AMSnet-q: Unsupervised Circuit Identification and Performance Labeling for AMS Circuits}

\author{Ze Zhang\textsuperscript{†,1}, Junzhuo Zhou\textsuperscript{†,2}, Yichen Shi\textsuperscript{4}, Zhuofu Tao\textsuperscript{2}, Rui Ji\textsuperscript{1}
\\[2pt]
Zhiping Yu\textsuperscript{*,3}, Quan Chen\textsuperscript{*,1}, Ting-Jung Lin\textsuperscript{*,4}, Lei He\textsuperscript{*,2}
\\[4pt]
\textsuperscript{1}Southern University of Science and Technology, \textsuperscript{2}University of California Los Angeles\\
\textsuperscript{3}Tsinghua University, \textsuperscript{4}Eastern Institute of Technology Ningbo\\[2pt]
\textsuperscript{†}Co-first authors \quad \textsuperscript{*}Corresponding authors
}

\maketitle

\begin{abstract}


Analog and mixed-signal (AMS) circuit design remains heavily reliant on expert knowledge. While recent AI-driven automation tools can generate candidate topologies, they critically depend on manually curated datasets with functional and performance annotations---a requirement that current large language models (LLMs) and vision models cannot automate. Existing approaches still require domain experts to manually interpret circuit functionality.

We present AMSnet-q, a fully automated, unsupervised pipeline that eliminates human-in-the-loop annotation by converting schematic images directly into a labeled AMS circuit database. Unlike prior work that stops at netlist extraction, our framework automates the complete verification loop: it performs schematic-to-netlist conversion, topology-aware testbench generation, and simulation-based sizing validation to objectively determine circuit functionality. Validated in 28 nm technology, AMSnet-q processed 739 schematics from the AMSnet 1.0 dataset, automatically constructing a repository of 4 circuit classes, 105 distinct topologies, and 89,789 labeled device configurations. By decoupling human effort from dataset volume and reducing the workload to a one-time testbench template per circuit class, AMSnet-q enables scalable, objective, and fully automated AMS database construction.

\end{abstract}






\section{Introduction}

AMS circuit design has traditionally been experience-driven, with engineers relying on tacit knowledge for topology selection and device sizing. Recent advances in artificial intelligence (AI) have catalyzed automated topology generation through LLMs and generative models~\cite{fayazi2023angel,shi2024amsnetkg,lai2025analogcoder}. Another category involves human-in-the-loop topology selection guided by algorithmic decision support ~\cite{Degrauwe1987IDAC,Khalil2022FastTopologySelection,Gerlach2017generictopologyselection}. However, these data-driven paradigms presuppose access to repositories of analog topologies annotated with trustworthy functionality and performance labels---a prerequisite that remains challenging to satisfy automatically.


The critical unmet challenge lies in functional identification and performance validation. Neither LLMs nor circuit-specific vision models can reliably interpret the functional semantics of a raw netlist (e.g., distinguishing a folded-cascode OpAmp from a comparator) or verify its electrical performance without human intervention. Consequently, experts must manually inspect each topology, configure type-specific testbenches, execute simulations, analyze results, and assign labels—a process whose effort grows linearly with dataset size and introduces subjective biases. 

Prior schematic recognition pipelines~\cite{tao2024amsnet,bhandari2024masala} only automate the visual recognition layer; the extracted netlists still demand labor-intensive manual annotation, deferring rather than alleviating the scalability bottleneck. Similarly, prior graph-based approaches\cite{wang2020gcn}, \cite{lee2025dice} are capable of encoding simple building blocks (e.g., differential pairs) through learned graph representations; however, these methods remain limited to implicit local relationships and struggle to achieve comprehensive functional understanding across the complete analog circuit.

\begin{figure}[!t]
    \centering
    \includegraphics[width=\linewidth]{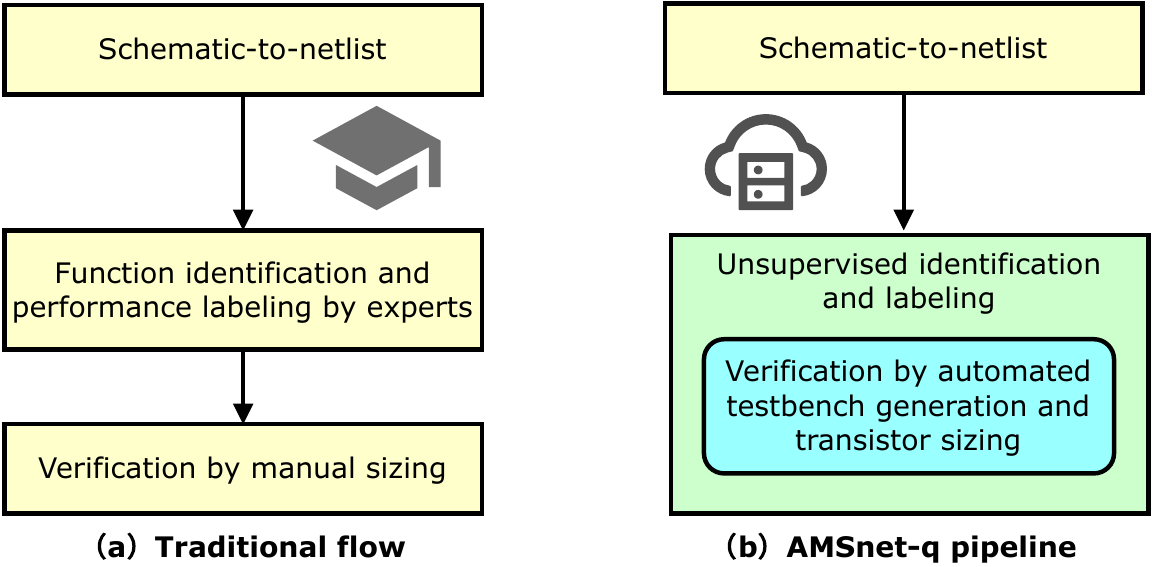}
    \caption{Traditional flow versus our proposed flow}
    \label{fig:overview}
\end{figure}

We propose AMSnet-q, an unsupervised framework that closes the automation gap by replacing manual annotation with executable simulation verification. As depicted in Fig.~\ref{fig:overview}, AMSnet-q completes the loop by integrating template-based testbench assembling with subsequent sizing-driven verification. Given a raw netlist produced by prior recognition pipeline, AMSnet-q first assembles a type-specific testbench from a reusable template library---constructed once per circuit class (\textit{e.g.}, OpAmp, Low-Dropout Regulator (LDO), comparator). This assembling ensures the executable simulation environment necessary for algorithmic sizing. Subsequently, an automated sizing engine explores the design parameter space to discover feasible device configurations that satisfy the requisite performance specifications under the assembled testbench. Only upon successful sizing and simulation does the framework determine circuit functionality and generate performance labels. 

AMSnet-q shifts the labor-intensive annotation into lightweight, one-time testbench template authoring per circuit type. This enables truly scalable AMS database construction without human intervention.
Our contributions are summarized as follows:

\begin{figure*}[!t]
    \centering
     \includegraphics[width=\textwidth]{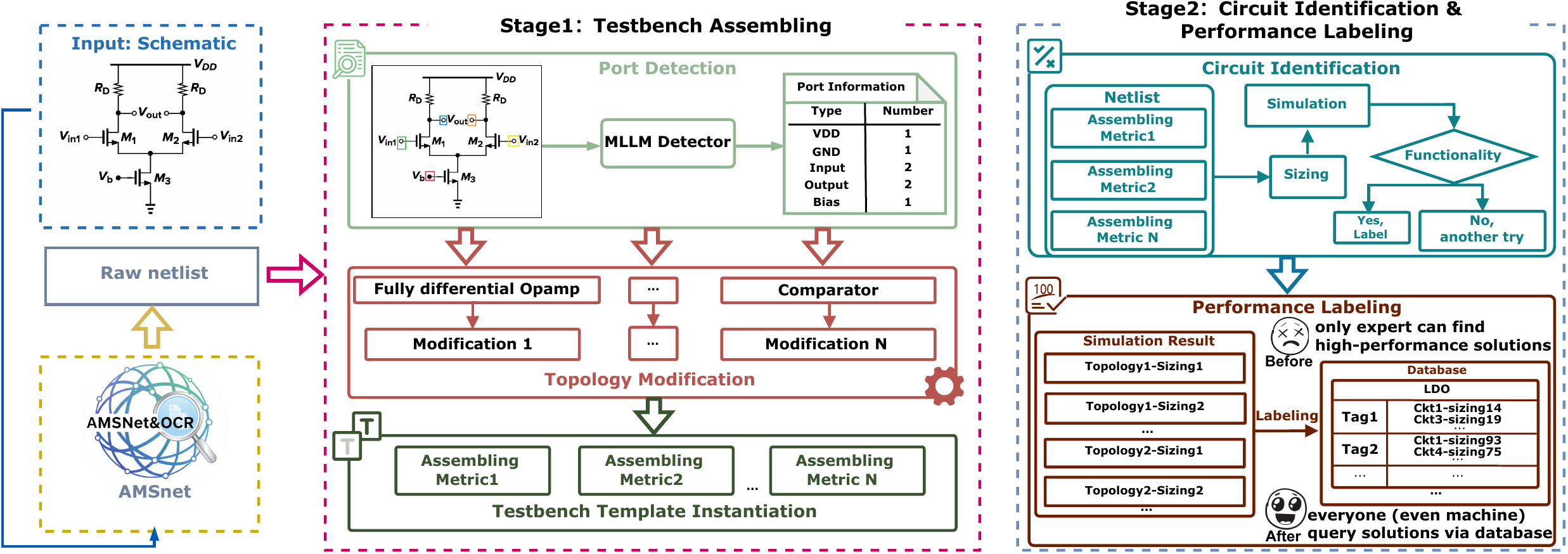}
    \caption{Proposed AMSnet-q Pipeline.}
    \label{fig:flow} 
\end{figure*}

\begin{itemize}
\item \textbf{Elimination of manual annotation through automated verification.} We replace subjective human labeling with objective simulation-based validation, automatically constructing a structured database including topology, functional class, and performance labels from raw schematic images.
\item \textbf{Topology-driven automated testbench assembly.} Given arbitrary netlists, the framework automatically detects ports and assembles executable test environments via intrusive topology modification, enabling unsupervised performance extraction.
\item \textbf{Scalable unsupervised dataset construction.} Verified with 739 schematics yielding 89,789 labeled configurations, demonstrating the feasibility of building large-scale AMS datasets without expert-in-the-loop curation.
\end{itemize}

\section{Preliminary}
\subsection{Schematic-to-Netlist Recognition}\label{sec:amsnet}
Schematic-to-netlist recognition automates transcription of published schematics into machine-readable netlists, as demonstrated at scale by AMSnet~\cite{tao2024amsnet,shi2025amsnet} and Masala-CHAI~\cite{bhandari2024masala}. However, these pipelines are confined to low-level geometry and cannot automatically infer port function (\textit{e.g.}, inputs/outputs, supply rails, bias, etc.) or circuit function, requiring expert inspection and correction. Consequently, simulation-ready netlists still require manual port labeling and functional interpretation before verification. While reducing structural netlist collection costs, these methods do not significantly decrease marginal human effort to produce functionally verified, testbench-ready designs.

\subsection{Automated Sizing}
\label{sec: auto-sizing}

Automated sizing aims to replace the manual tuning of device dimensions and bias conditions. The engine iteratively proposes parameters, invoking SPICE simulation, and evaluating performance against specifications. Methods spanning Bayesian optimization (BO)~\cite{mockus2012bayesian,Lyu2018bo,chen2022gmid} with surrogate models, evolutionary and genetic algorithms~\cite{nsga-ii,electronics10243148}, reinforcement learning~\cite{Zhao2020DRL,Budak2021dnnopt}, and others, converge on a unified abstraction: the circuit-simulator loop constitutes an expensive black-box evaluator. This yields the constrained optimization problem:
\begin{equation}\label{eq:sizing}
  \min_{\bm{x} \in \mathcal{X}} \; f(\bm{x}) \quad \text{subject to} \quad g_m(\bm{x}) \geq 0,\; m = 1,\dots,M,
\end{equation}

\noindent
where $f$ and $g_m$ are accessible exclusively through simulation and are typically non-convex, noisy, and expensive to evaluate. At their core, these frameworks differ only in exploration strategy while functioning as unified black-box solvers.

\section{AMSnet-q Pipeline}

As illustrated in Fig.~\ref{fig:flow}, AMSnet-q comprises two successive stages that transform raw schematic netlists into a structured, labeled database without human intervention. Details of each stage are presented in following subsections.




\subsection{Stage 1: Testbench Assembling}

AMSnet-q starts with the image-to-netlist pipeline. This stage aims to recover semantic information from recognition result and constructs executable simulation environments. Given a netlist with unlabeled port symbols, AMSnet-q performs three steps, \textbf{Port-type Detection, Topology Modification, and Testbench Instantiation}, to produce simulation-ready decks. 

\textbf{Port-type Detection:} 
Although schematic image recognition extracts netlists retaining geometric port symbols, the absence of semantic labels (\texttt{VIN}, \texttt{VOUT}, \texttt{VBIAS}) precludes automated testbench assembly. To interpret these visual cues without manual inspection, AMSnet-q harnesses the visual comprehension capabilities of a multimodal large language model (MLLM) via tailored visual prompting. 

\begin{figure}[!t]
    \centering
    \includegraphics[width=\linewidth]{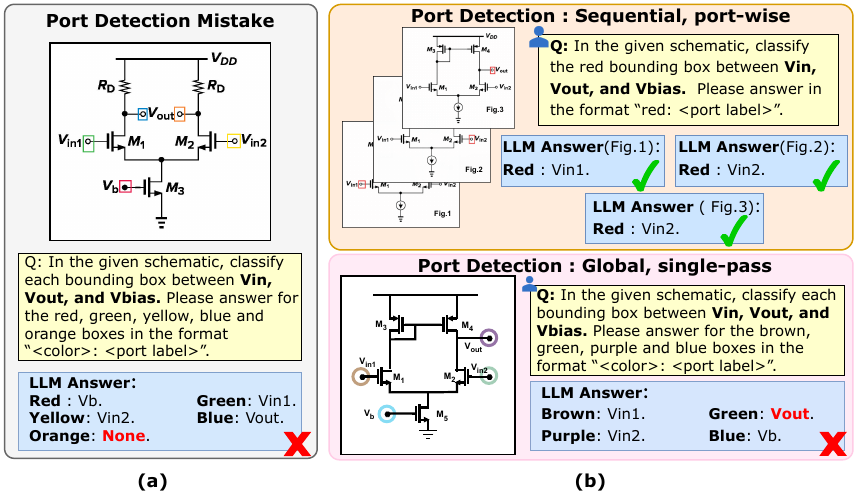}
    \caption{Illustration of port detection: (a) a mistake example and (b) two strategies for port detection.}
    \label{fig:port_classification_mistake_new}
\end{figure}

Specifically, we compare two querying strategies (Fig.~\ref{fig:port_classification_mistake_new}): (i)~\emph{global single-pass}, which visualizes all ports simultaneously, versus (ii)~\emph{sequential port-wise}, which isolates each port through individual visual prompts. The single-pass approach suffers from color confusion and cross-port interference, whereas sequential querying eliminates contextual ambiguity, yielding higher accuracy ($94.22\%$ vs. $69.55\%$).
AMSnet-q therefore adopts the port-wise strategy.

\begin{figure}[!t]
    \centering
    \includegraphics[width=0.60\linewidth]{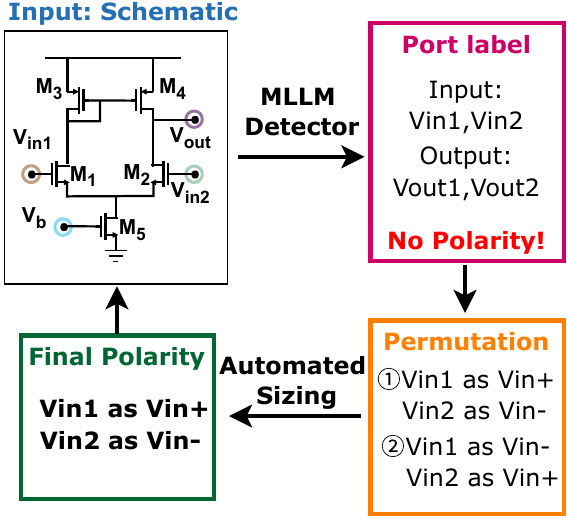}
    \caption{Illustration of port permutation.}
    \label{fig:permutation_flow}
\end{figure}

After port detection resolves port types, AMSnet-q addresses polarity ambiguity (\textit{e.g.}, \texttt{Vin} lacking explicit ``$+$/$-$'' marks). Rather than relying on naming heuristics, Fig.~\ref{fig:permutation_flow} shows the framework that (1) identifies potential differential pairs via structural priors, (2) enumerates the constrained set of polarity permutations, and (3) validates each candidate through automated sizing simulations. The permutation achieving functional specifications with minimal optimization effort is adopted as the ground-truth configuration. This establishes a closed verification loop---\emph{port detection} $\rightarrow$ \emph{permutation enumeration} $\rightarrow$ \emph{simulation validation}---that augments the DUT netlist with verified functional port annotations, shown as Fig.~\ref{fig:permutation_flow}.

\textbf{Topology Modification:}
Port detection yields a labeled netlist, yet raw topologies often lack the access points required for performance evaluation. AMSnet-q performs intrusive topology modification to expose instrumentation nodes while preserving the operating point and external I/O behavior. The modifications are executed automatically based solely on port specifications and connectivity, without manual intervention.

\begin{figure}[!t]
    \centering
    \includegraphics[width=1\linewidth]{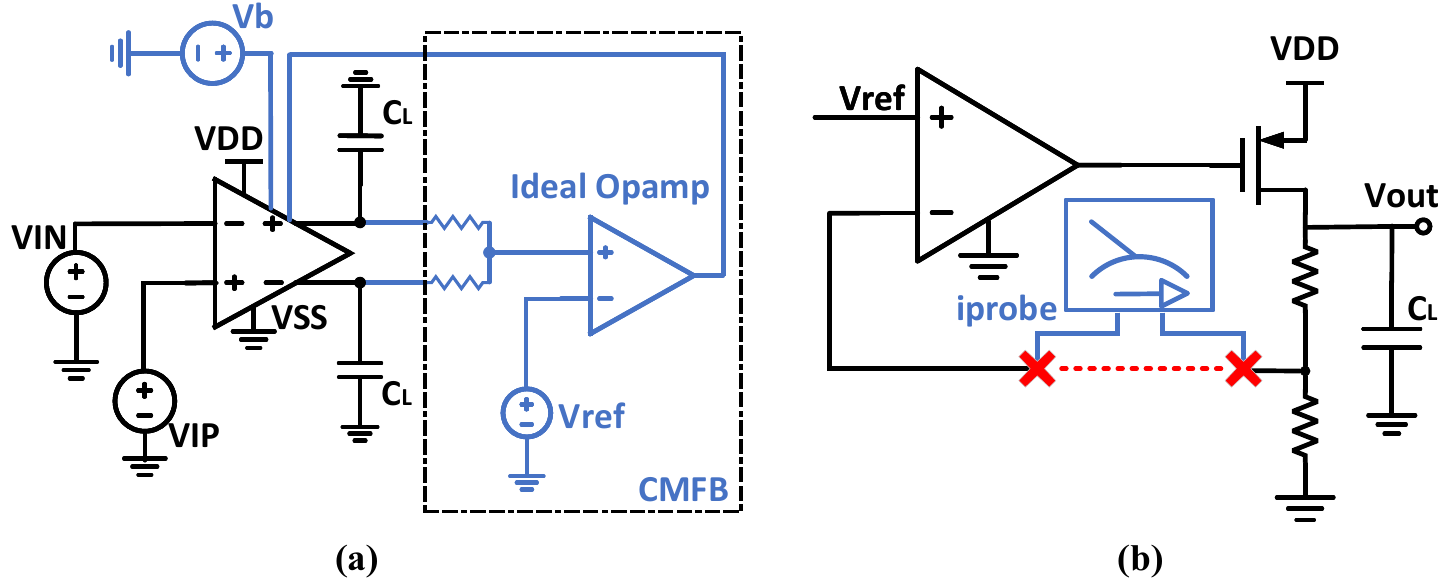}
    \caption{Topology modification: (a) Fully-Diff. OpAmp and (b) LDO.}
    \label{fig:LDO case-study}
\end{figure}

Fig.~\ref{fig:LDO case-study} presents two examples of DUT topology modification. For Fully-Differential OpAmps (Fig.~\ref{fig:LDO case-study}a), the modification necessity hinges on output stage characteristics. High-impedance current-source loads leave the output common-mode level electrically floating without a DC feedback path, necessitating explicit Common-Mode Feedback (CMFB) injection to stabilize the operating point. Conversely, resistive loads or structural dividers intrinsically fix the common-mode level, often obviating additional CMFB. The framework discriminates these scenarios via port signature analysis, applying either explicit CMFB attachment or harness-mode actuation of existing bias nodes accordingly.

For LDOs (Fig.~\ref{fig:LDO case-study}b), AMSnet-q identifies the pass-transistor and feedback divider, then severs the connection between the divider output and error amplifier input to insert an iprobe. This probe insertion is essential to characterize the internal error amplifier's performance metrics (\textit{e.g.}, loop gain, phase margin) while preserving the circuit's DC stability during simulation.

\textbf{Testbench Instantiation:} Once topology modification is complete, AMSnet-q assembles a reusable, parametric testbench template around the modified DUT. Each template specifies a functional class (e.g., OpAmp, LDO), measurable metrics, and required port signatures (number, type, polarity). Selection proceeds via rule-based filtering for port coverage and metric compatibility; when multiple candidates exist, the system instantiates them in parallel for subsequent validation.

Crucially, templates remain \emph{bias-agnostic} to accommodate design-specific variability. Rather than hard-coding bias sources, the framework identifies bias terminals dynamically and inserts corresponding modules with sensible initial ranges on demand, allowing the sizing engine to tune operating points without jeopardizing convergence.

Instantiation then proceeds by:
\begin{enumerate}
\item wiring template sources and measurement blocks to annotated DUT ports;
\item attaching type-specific harnesses (e.g., CMFB, LDO iprobe) produced during topology modification;
\item emitting a fully specified SPICE deck with stimulus and measurement directives.
\end{enumerate}

At this point, AMSnet-q has transformed schematic-derived netlists into simulation-ready decks. This operational paradigm shifts the burden from $O(N)$ manual labeling (where $N$ is the number of topologies) to $O(1)$ testbench template authoring per circuit class, enabling scalable AMS database construction.

\begin{algorithm}[t]
  \caption{Unsupervised Performance Labeling}
  \label{alg:perf-label}
  \KwIn{Score matrices $\{\mathbf{S}_c\}$ for all topologies of a circuit type; number of clusters $K$.}
  \KwOut{Label $\ell_m$ and textual tag $\mathcal{T}^{(k)}$ for each trial.}

  Stack $\{\mathbf{S}_c\}$ into a global score matrix $\mathbf{S}\in\mathbb{R}^{M\times D}$\;
  Compute global mean $\bm{\mu}$ and standard deviation $\bm{\sigma}$ over $\mathbf{S}$\;
  Standardize scores: $\mathbf{Z} \leftarrow (\mathbf{S}-\bm{\mu})/\bm{\sigma}$\;
  Truncate negatives: $\mathbf{Z}_+ \leftarrow \max(\mathbf{Z}, 0)$\;
  Run $K$-means on $\mathbf{Z}_+$ to obtain labels $\ell_1,\dots,\ell_M$ and cluster memberships\;
  Compute cluster centers in score space
  $\bar{\bm{s}}^{(k)}$ by averaging rows of $\mathbf{S}$ within each cluster $k$\;
  \For{each metric $j=1,\dots,D$}{
    Collect $\{\bar{s}^{(k)}_j\}_{k=1}^K$ and sort clusters by this value\;
    Mark top (bottom) fraction of clusters as \texttt{good} (\texttt{bad}) for metric $j$; mark the rest as \texttt{moderate}\;
  }
  \For{each cluster $k=1,\dots,K$}{
    Compute standardized center
    $\bar{\bm{z}}^{(k)} \leftarrow (\bar{\bm{s}}^{(k)}-\bm{\mu})/\bm{\sigma}$\;
    Rank metric indices $j$ by $|\bar{z}^{(k)}_j|$ in descending order\;
    Select the first few metrics and form a textual tag
    $\mathcal{T}^{(k)}$ from their (\texttt{good}/\texttt{moderate}/\texttt{bad}, metric\_name) pairs\;
  }
  Assign to each trial $m$ the performance label $(\ell_m, \mathcal{T}^{(\ell_m)})$\;

\end{algorithm}


\begin{figure*}[!t]
    \centering
    \includegraphics[width=\textwidth]{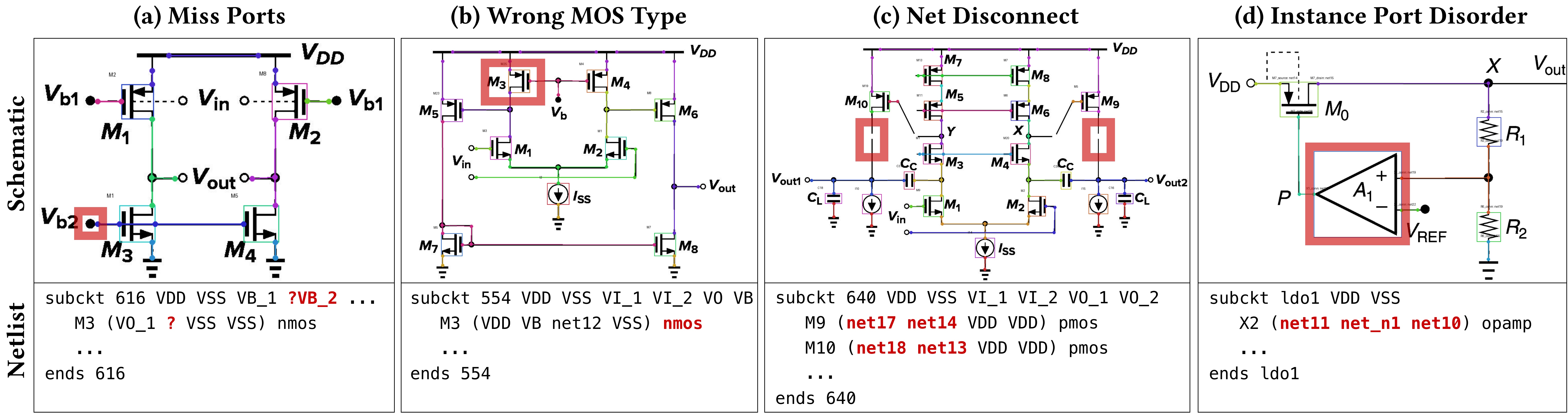}
    \caption{False topologies identified by AMSnet-q circuit identification. \scriptsize{Sources:  B.Razavi~\cite{razavi2001design} (a) page 408, (b) page 362, (c) page 439; (d) \cite{Razavi2022LDO} page 8.}}
    \label{fig:failure study}
\end{figure*}

\subsection{Circuit Identification and Performance Labeling}

Leveraging the assembled testbenches from stage 1, stage 2 operationally validates circuit functionality and generates performance labels. First, the \textbf{Operational Circuit Identification} step relies on an automated sizing engine to explore device parameter spaces to determine whether a topology satisfies functional specifications. A circuit is classified as type $C$ if and only if feasible parameters exist. In the \textbf{Unsupervised Performance Labeling} step, validated topologies undergo extended optimization, with resulting performance vectors clustered into human-readable trade-off labels.

\textbf{Operational Circuit Identification:}
Unlike prior work that relies on LLMs or human experts to infer circuit functionality, AMSnet-q treats classification as a feasibility problem. A topology is identified as candidate type $C$ if and only if there exists a parameter vector $\boldsymbol{x} \in \mathcal{X}$ satisfying that type's threshold specifications $\mathcal{R}_m$:
\begin{equation}
\text{find } \boldsymbol{x} \in \mathcal{X} \quad \text{s.t.} \quad g_{m}(\boldsymbol{x})\geq \mathcal{R}_{m}, \quad m=1,\ldots,M
\end{equation}
where $g_m(\cdot)$ denotes the $m$-th simulated performance metric. If the automated sizing engine exhausts its trial budget $B$ without discovering feasible $\boldsymbol{x}$, the topology is rejected for type $C$; otherwise, it is accepted and forwarded for performance labeling. This operational definition eliminates subjective biases inherent in manual annotation.

For topologies with ambiguous port signatures, multiple candidate types are proposed and validated in parallel. The framework automatically corrects netlist recognition errors (Fig.~\ref{fig:failure study}) through this simulation-based consistency check, reducing false negatives to zero upon fixing structural inaccuracies.

\textbf{Unsupervised Performance Labeling:}
Once validated, topologies undergo extended sizing with higher performance targets $\mathcal{R}_m^* > \mathcal{R}_m$ and increased trial budget $B^* > B$. Each successful trial yields a normalized score vector $\boldsymbol{s} \in [0,100]^M$, where meeting the identification threshold $\mathcal{R}_m$ corresponds to a score of 60 and achieving the target $\mathcal{R}_m^*$ yields 100.

To discover recurring trade-off patterns, we aggregate performance score vectors into matrix $\mathbf{S} \in \mathbb{R}^{N \times M}$ ($N$ topologies, $M$ metrics). Following Algorithm~\ref{alg:perf-label}, we standardize $\mathbf{S}$ via Z-score normalization to obtain $\mathbf{Z} = (\mathbf{S} - \bm{\mu}) / \bm{\sigma}$, then truncate negative entries to form $\mathbf{Z}_+ = \max(\mathbf{Z}, 0)$, thereby focusing clustering on performance strengths rather than deficits. Applying $K$-means~\cite{KMeans} to $\mathbf{Z}_+$ yields cluster assignments $\ell_n$ and centroids $\bar{\bm{s}}^{(k)}$ in the original score space.

For each cluster $k$, we rank centroid values $\bar{s}^{(k)}_j$ relative to other clusters; top/bottom quantiles are rated \texttt{good}/\texttt{bad} for metric $j$, with intermediates marked \texttt{moderate}. We then rank metrics by the magnitude of their standardized centroid deviation $|\bar{z}^{(k)}_j|$ and retain the top 3--4 salient features to compose concise human-readable trade-off tags (e.g., ``good Gain; bad Area''). Each trial inherits its cluster index and corresponding tag, constituting the final unsupervised performance labels.


\begin{figure*}[!t]
    \centering
    \includegraphics[width=\linewidth]{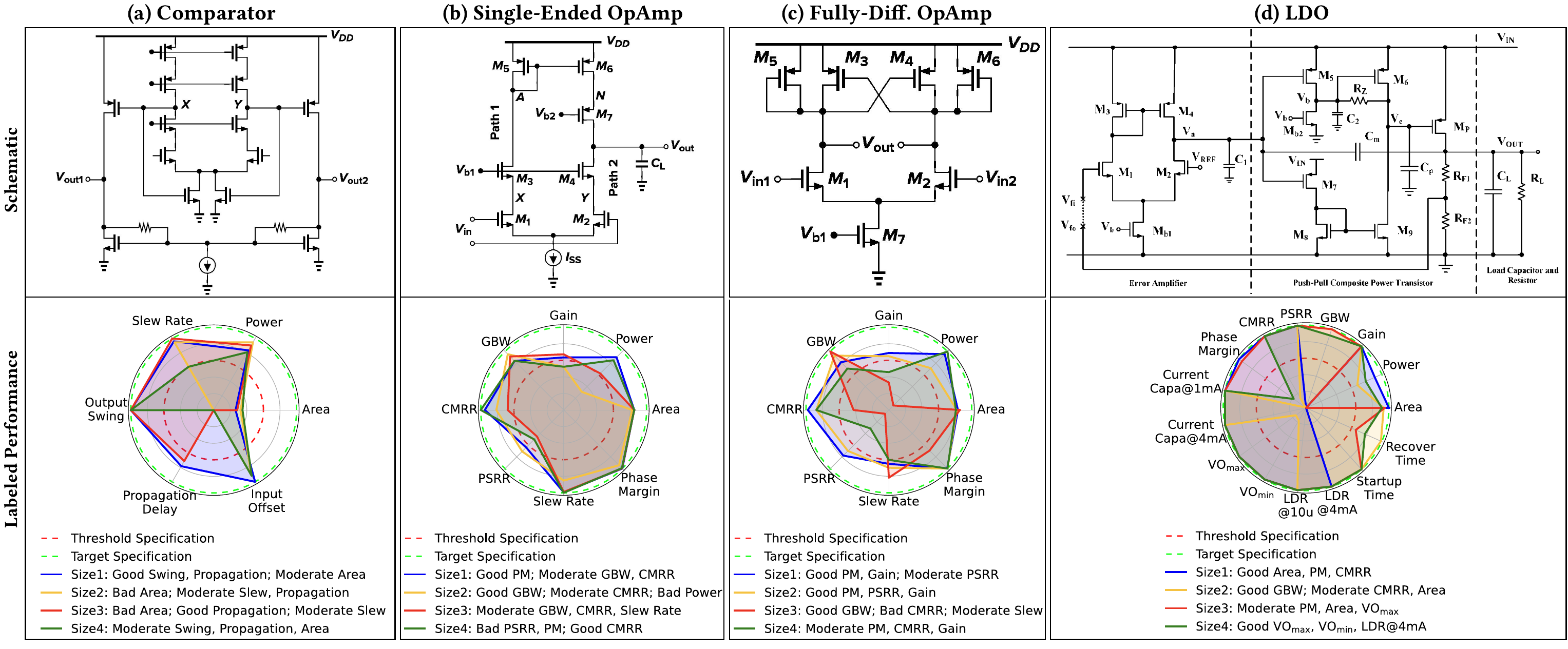}
    \caption{Performance labeling and database example. \scriptsize{Sources: B.Razavi~\cite{razavi2001design} (a) page 389, (b) page 421, (c) page 131; (d) G.Jaswanth~\cite{Jaswanth2015ACL} page 6.}}
    \vspace{-5mm}
    \label{fig:performance labeling result}
\end{figure*}

\section{Experiment}

\begin{table}[!t]
\centering
\caption{Threshold and Target Specs used in experiments.}
\label{tab:spec}
\resizebox{\columnwidth}{!}{
\begin{tabular}{|c|c|c|c|c|}
\hline
\textbf{Circuit} & \multirow{2}{*}{\textbf{Metric}} & \multirow{2}{*}{\textbf{Unit}} & \textbf{Threshold}$^*$ & \textbf{Target} $^\dagger$ \\
\textbf{Type} & & & \textbf{Specification} & \textbf{Specification} \\
\hline\hline
\multirow{8}{*}{OpAmp}
 & Power        & $\mu$W      & 4.5   & 0.5   \\
 & Gain         & dB      & \cellcolor[HTML]{fef9e5} 40    & 80    \\
 & GBW          & kHz     & \cellcolor[HTML]{fef9e5} 100   & 10000 \\
 & Phase Margin & degree  & \cellcolor[HTML]{fef9e5} 30    & 45    \\
 & Slew Rate    & V/$\mu$s    & 0.1   & 8000  \\
 & CMRR         & dB      & 40    & 90    \\
 & PSRR         & dB      & -40   & -60   \\
 & Area         & $\mu$m$^2$  & 8     & 1     \\
\hline
\multirow{6}{*}{Comp.}
 & Power            & $\mu$W     & 100   & 20    \\
 & Output Swing Voltage         & V      & \cellcolor[HTML]{fef9e5}$\mathrm{VDD}/2$ & $2\mathrm{VDD}/3$ \\
 & Slew Rate        & V/$\mu$s   & 100   & 2000  \\
 & Propagation delay& ns     & 0.8   & 0.2   \\
 & Input Offset Voltage (3$\sigma$) & mV     & 10    & 1     \\
 & Area             & $\mu$m$^2$ & 4     & 1     \\
\hline
\multirow{14}{*}{LDO}
 & Power ($98^\circ C$)         & $\mu$W     & 14.4  & 10.8  \\
 & VO                           & V      & \cellcolor[HTML]{fef9e5}(1.5975, 1.6025) & (1.5985, 1.6015) \\
 & Current Capa. (1mA load)     & mV     & \cellcolor[HTML]{fef9e5}0.9   & 1     \\
 & Current Capa. (4mA load)     & mV     & \cellcolor[HTML]{fef9e5}3.6   & 4     \\
 & Load Regulation (1uA load)   & mV     & \cellcolor[HTML]{fef9e5}0.9   & 1     \\
 & Load Regulation (4mA load)   & mV     & \cellcolor[HTML]{fef9e5}27    & 30    \\
 & Gain                         & dB     & 40    & 45    \\
 & GBW                          & kHz    & 90    & 180   \\
 & Phase Margin                 & degree & 30    & 45    \\
 & CMRR ($98^\circ C$)          & dB     & 55    & 60    \\
 & PSRR ($98^\circ C$)          & dB     & -36   & -40   \\
 & Startup Time                 & $\mu$s     & 30    & 20    \\
 & Recovery Time                & ns     & 2000  & 800   \\
 & Area                         & $\mu$m$^2$ & 60    & 20    \\
\hline
\end{tabular}}
\parbox{\linewidth}{\footnotesize
$^*$ \colorbox[HTML]{fef9e5}{Those thresholds} serve as $\mathcal{R}_m$ in circuit identification, other thresholds are only used to obtain normalized scores.
$^\dagger$ Serve as $\mathcal{R}_m^*$ in performance labeling.
}
\end{table}

We validate AMSnet-q on 739 schematics from the AMSnet dataset~\cite{tao2024amsnet}. The experimental evaluation proceeds in two stages: we first assess the accuracy of circuit identification and the quality of performance labeling independently, then demonstrate the scalability of the complete pipeline through comprehensive database construction. 

The pipeline employs Gemini-3.1-pro~\cite{google2025gemini3pro} for port detection, NSGA-II~\cite{nsga-ii,akiba2019optuna} for automated sizing, and Cadence SPECTRE for circuit simulation, targeting TSMC 28nm technology at the \texttt{TT, 25°C} corner. Table~\ref{tab:spec} summarizes the performance specifications governing circuit identification and labeling. We configure trial budgets of $\mathcal{B}=200$ for feasibility identification and $\mathcal{B}^*=2000$ for performance optimization.




\subsection{Validation of Circuit Identification}
\begin{figure*}[!t]
    \centering
    \vspace{-6mm}
    \includegraphics[width=\linewidth]{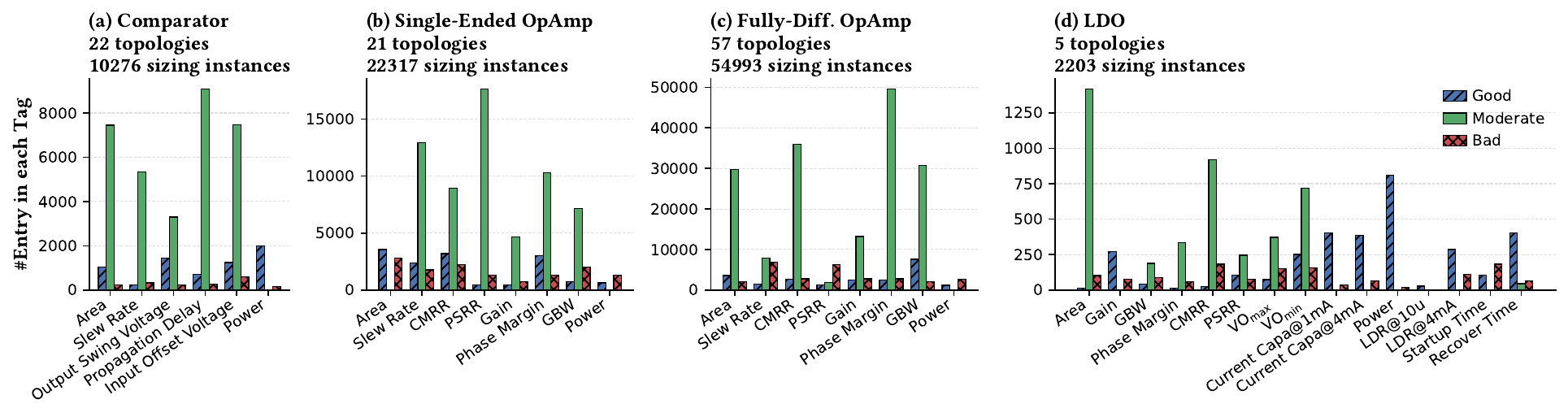}
    \vspace{-10mm}
    \caption{Summary of AMSnet-q database.}
    \vspace{-3mm}
    \label{fig:summary}
\end{figure*}

We evaluate circuit identification accuracy on 734 schematics from the AMSnet dataset, among which we manually annotate 22 Single-Ended and 66 Fully-Differential OpAmps as ground truth. Following the operational definition in  Table\ref{tab:circuit-identify}, we classify a schematic as an OpAmp only if it satisfies minimum performance thresholds: 40\,dB DC gain, 100\,kHz gain-bandwidth product (GBW), and $30^{\circ}$ phase margin (PM).

AMSnet-q automatically identifies 15 Single-Ended and 47 Fully-Differential OpAmps with correct port permutations, achieving F1 scores of 0.811 and 0.832. The verification process consumes 20{,}750 simulations (6.4 hours on a 16-core CPU). This cost is dominated by exhaustive polarity enumeration (2$\times$ permutations for Single-Ended, 4$\times$ for Fully-Differential) and negative cases that exhaust the trial budget $\mathcal{B}$ without meeting specifications. Notably, the only manual effort required is the one-time construction of testbench templates.

Representative false negative (FN) cases---topologies incorrectly rejected despite being valid OpAmps---are illustrated in Fig.~\ref{fig:failure study}. These errors primarily stem from netlist recognition inaccuracies (e.g., missing ports, disconnected nets) rather than identification algorithm failures. Upon manual correction of these structural defects, the FN rate drops to zero, improving F1 scores to 0.977 (Single-Ended) and 0.927 (Fully-Differential). This demonstrates AMSnet-q's capability to robustly validate circuit topology and effectively detect inconsistencies in upstream recognition results.

\begin{table}[!t]
\centering
\captionsetup{justification=centering}
\caption{Identification confusion matrix for OpAmp.
}
\label{tab:circuit-identify}


\footnotesize
\begin{tabular}{lccc}
\toprule
\diagbox[width=2cm,height=0.55cm]{\textbf{Actual}}{\textbf{Pred.}} & \textbf{Single-Ended} & \textbf{non Single-Ended} & \textbf{Total} \\
\midrule
\textbf{Single-Ended}     & \cellcolor[HTML]{ebfaed} TP = 15 $\to$ 21 $^\dagger$  & \cellcolor[HTML]{fce3e4}FN = 6  $\to$ 0 $^\dagger$& 21  \\
\textbf{non Single-Ended} & \cellcolor[HTML]{fce3e4}FP = 1  & \cellcolor[HTML]{ebfaed} TN = 712 & 713 \\
\midrule
\textbf{Total}    & 16      & 718      & 734 \\
\bottomrule \\
\toprule
\diagbox[width=2cm,height=0.55cm]{\textbf{Actual}}{\textbf{Pred.}} & \textbf{Full-Diff.} & \textbf{non Full-Diff.} & \textbf{Total} \\
\midrule
\textbf{Full-Diff.}     &  \cellcolor[HTML]{ebfaed}TP = 47 $\to$ 57 $^\dagger$  & \cellcolor[HTML]{fce3e4} FN = 10  $\to$ 0 $^\dagger$ & 57  \\
\textbf{non Full-Diff.} & \cellcolor[HTML]{fce3e4}FP = 9  & \cellcolor[HTML]{ebfaed}TN = 668 & 677 \\
\midrule
\textbf{Total}    & 56      & 678      & 734 \\
\bottomrule
\end{tabular}
\begin{minipage}{\columnwidth}
\footnotesize
\colorbox[HTML]{ebfaed}{\parbox{0.3cm}{TP}}: A function supported by the schematic has been correctly identified; \\[-1mm]
\colorbox[HTML]{ebfaed}{\parbox{0.3cm}{TN}}: A function unsupported by the schematic has been correctly disqualified; \\[-1mm]
\colorbox[HTML]{fce3e4}{\parbox{0.3cm}{FP}}: A function unsupported by the schematic has been incorrectly identified;\\[-1mm]
\colorbox[HTML]{fce3e4}{\parbox{0.3cm}{FN}}: A function supported by the schematic has been incorrectly disqualified.\\[-1mm]
\quad\quad\quad $^\dagger$ AMSnet-a successfully figures out those incorrect topologies, as in Fig~\ref{fig:failure study}.
\end{minipage}
\end{table}

\subsection{Validation of  Performance Labeling}

\label{sec:exp-perf-labeling}

To assess the quality of performance labels, we examine whether discovered clusters capture meaningful design trade-offs and whether generated tags provide compact, interpretable summaries of achievable performance regions. We set $K = 30$ clusters for $K$-means partitioning across all circuit types.

Fig.~\ref{fig:performance labeling result}(b) illustrates the resulting labels for a Single-Ended OpAmp topology. The radar charts compare two representative sizing configurations: ``Size1'' (blue) exhibits superior phase margin (PM) with moderate gain-bandwidth product (GBW) and common-mode rejection ratio (CMRR), while ``Size2'' (yellow) achieves higher GBW at the cost of increased power consumption. These automatically generated tags enable designers to rapidly identify Pareto-optimal configurations matching specific requirements---for instance, selecting ``Size1'' for stability-critical applications or ``Size2'' for high-speed scenarios---without manual inspection of raw simulation data. 

Similar trade-off visualizations are presented in Fig.~\ref{fig:performance labeling result}(a, c, d), demonstrating that the clustering consistently reveals physically meaningful performance correlations across diverse analog circuit classes. The radar chart overlays two reference boundaries: the red (score of 60) and green (score of 100) dashed lines mark the Threshold Specification and Target Specification in Table ~\ref{tab:spec}, respectively.


\subsection{AMSnet-q Database Construction}

Upon validating the identification and labeling methodologies, we construct the AMSnet-q database by processing all experimental results through the complete pipeline. The performance tags enable efficient retrieval of topologies and sizing configurations without parsing raw simulation data, significantly streamlining design-space exploration. 

The resulting repository comprises 4 circuit categories (Single-Ended OpAmps, Fully-Differential OpAmps, comparators, and LDOs), 105 distinct topologies, and 89,789 labeled device-sizing instances. This structured dataset provides a comprehensive foundation for data-driven analog circuit design research. Fig.~\ref{fig:summary} illustrates the distribution of entries across the four circuit types and their performance classifications.

\section{Conclusion}

We present AMSnet-q, a scalable unsupervised framework that closes the automation loop for AMS circuit database construction~\cite{patent_ams_2025}. By replacing manual functional annotation with executable simulation verification, it reduces human effort from $O(N)$ per-topology labeling to $O(1)$ testbench template design per circuit class. Integrating MLLM-based visual understanding, intrusive topology modification, and automated sizing, the pipeline transforms raw schematic images into structured netlists with operational classifications and quantitative performance labels. Validated in TSMC 28nm technology on 739 schematics, AMSnet-q constructed a dataset comprising 4 circuit categories, 105 topologies, and 89,789 labeled configurations, providing a robust foundation for data-driven analog design. Future work will extend to complex mixed-signal systems, enable cross-technology portability, and integrate with automated layout synthesis for a fully automated schematic-to-layout workflow.

\end{CJK*}


\end{document}